\documentstyle[twoside,fleqn,jhuwkshp,psfig]{article}
\newcommand{\etal}{{\it et al.} }

\newcommand{\xmm}{{\it XMM-Newton} }
\newcommand{\chandra}{{\it Chandra} }

\newcommand{\astroep}{{\it Astro-E2}}
\newcommand{\rxte}{{\it RXTE} }
\newcommand{\rxtep}{{\it RXTE}}
\newcommand{\hetg}{{\it HETGS} }
\newcommand{\hetgp}{{\it HETGS}}

\newcommand{\fekalfa}{{Fe~K$\alpha$} }

\newcommand{\fekbeta}{{Fe~K$\beta$} }

\newcommand{\fexxv}{Fe~{\sc xxv} }

\newcommand{\feklya}{{Fe~{\sc xxvi }~Ly$\alpha$} }

\newcommand{\feklyap}{{Fe~{\sc xxvi}~Ly$\alpha$}}
\newcommand{\resonetwo}{{$1s^{2}-1s2p$} }

\newcommand{\bsax}{{\it BeppoSAX} }
\newcommand{\figlc}{{Fig.~1} }
\newcommand{\figlcp}{{Fig.~1}}
\newcommand{\figcontmodels}{{Fig.~3} }
\newcommand{\figcontmodelsp}{{Fig.~3}}

\newcommand{\figcontsnapshots}{{Fig.~4} }
\newcommand{\figcontsnapshotsp}{{Fig.~4}}
\newcommand{\figtwofekrat}{{Fig.~5} }

\newcommand{\figsixspecrat}{{Fig.~2} }

\newcommand{\figcompshoulder}{{Fig. 6} }
\newcommand{\figcompshoulderp}{{Fig. 6}}

\newcommand{\tablefits}{Table~1 }
\newcommand{\tablefitsp}{Table~1}

\newcommand{\src}{NGC~3783 }

\begin{document}

\title{\chandra HIGH-RESOLUTION X-RAY SPECTROSCOPY OF THE Fe~K LINE 
IN THE SEYFERT 1 GALAXY NGC~3783}

\author{T. Yaqoob\address{\it Department of Physics and Astronomy,
Johns Hopkins University, Baltimore, MD 21218.}
\address{\it Laboratory for High Energy Astrophysics,
NASA/Goddard Space Flight Center, Greenbelt, MD 20771.},
J. N. Reeves$^{\rm a \ b}$ \address{\it Universities Space Research Association.},
A. Markowitz$^{\rm b}$ \address{\it N.A.S./N.R.C. Research Associate} ,
P. J. Serlemitsos$^{\rm b}$,
and U. Padmanabhan$^{\rm a}$ }

\begin{abstract}
\vspace{-3mm}
\centerline{\bf Abstract}

We report on the results of detailed X-ray spectroscopy
of the Fe-K region in the Seyfert~1 galaxy NGC~3783 using the \chandra
High Energy Grating Transmission Spectrometer (\hetgp).
There were five observations over an interval of $\sim 125$ days
in 2001,
each with an exposure time of $\sim 170$~ks. The
combined data constitute the highest signal-to-noise Fe-K
spectrum having the best velocity resolution in the Fe-K band to
date (FWHM $\sim 1860 \ \rm km \ s^{-1}$). The combined data
show a resolved \fekalfa line core (FWHM~$=1700^{+410}_{-390} \rm 
\ km \ s^{-1}$) with a center energy of $6.397 \pm 0.003$~keV,
consistent with an origin in neutral or lowly ionized Fe, 
located between
the BLR and NLR, as found by Kaspi \etal (2002). 
We also find that excess flux around the base
of the \fekalfa line core can be modeled with 
either a Compton scattering ``shoulder'' or an emission line
(with about the same flux as the line core) from a relativistic
accretion disk, having an inclination angle of $11^{\circ}$ or less.
This disk line model is as good as a 
Compton-shoulder model
for the base of the \fekalfa line core.
In the latter model, we measured the column density to be
$7.5^{+2.7}_{-0.6} \times \ \rm 10^{23} \ cm^{-2}$, which
corresponds to a Thomson optical depth of $\sim 0.60$, so the
line-emitting matter is not quite Compton-thick.
An intrinsic width of $1500^{+460}_{-340} \rm \ km \ s^{-1}$ FWHM
is still required in this model.
Moreover, more complicated 
scenarios involving both a Compton-shoulder and a disk
line cannot be ruled out. 
We confirm an absorption feature 
due to He-like Fe (FWHM~$=6405^{+5020}_{-2670} \rm \ km \ s^{-1}$), 
found in previous studies.


\centerline{\it To appear in the Astrophysical Journal, 10 July 2005}

\end{abstract}
\vspace{5mm}
\maketitle

\section{INTRODUCTION}\label{intro}

\begin{table*}[htb]
\begin{center}
\caption{Core Fe K Line $\it Chandra$ (HEG) Spectral Fitting Results}
\end{center}
\label{tab:npagetab}
{\small
\begin{tabular*}{\textwidth}{@{}l@{\extracolsep{\fill}}rrrrr}
& & & & & \\
\hline
& & & & & \\
Observation & {$E^{a}$} & {$I^{b}$} & {$  {\rm EW}^{c}$} & {$ {\rm FWHM}^{d}$} & {$ F^{e}$} \\
& (keV) & &  (eV) & {($\rm km \ s^{-1}$)} & {$ L^{e}$} \\
& & & & & \\
\hline
& & & & & \\
NGC 3783(1) & $6.401^{+0.008}_{-0.008}$ & $4.77^{+1.31}_{-1.17}$ & $75^{+21}_{-18}$ & $2320^{+1225}_{-995}$ & $5.0$ \\

   & ($6.390 - 6.412$) & ($3.22 - 6.58$) & ($51 - 103$) & ($990 - 4055$) & $1.1$ \\

NGC 3783(2) & $6.393^{+0.008}_{-0.009}$ & $5.42^{+1.39}_{-1.22}$ & $85^{+22}_{-19}$ & $2545^{+1300}_{-915}$ & $5.1$ \\

   & ($6.380 - 6.403$) & ($3.82 - 7.33$) & ($60 - 115$) & ($1365 - 4415$) & $1.1$ \\

NGC 3783(3) &  $6.394^{+0.005}_{-0.005}$ & $5.19^{+1.20}_{-1.08}$ & $81^{+19}_{-17}$ & $1205^{+940}_{-1205}$ & $5.2$ \\

& ($6.387 - 6.401$) & ($3.77 - 6.83$) & ($59 - 107$) & ($0 - 2540$) & $1.1$ \\

NGC 3783(4)  &  $6.396^{+0.005}_{-0.005}$ & $5.61^{+1.25}_{-1.15}$ & $66^{+15}_{-14}$ & $1255^{+760}_{-1255}$ & $7.4$ \\

   & ($6.389 - 6.402$) & ($4.08 - 7.31$) & ($48 - 86$) & ($0 - 2305$) & $1.6$  \\

NGC 3783(5) &   $6.401^{+0.007}_{-0.006}$ & $3.86^{+1.13}_{-1.02}$ & $51^{+15}_{-13}$ & $1345^{+990}_{-1345}$ & $6.1$ \\

   & ($6.392 - 6.411$) & ($2.52 - 5.39$) & ($33 - 71$) & ($0 - 2715$) & $1.3$ \\

Total &  $6.397^{+0.003}_{-0.003}$ & $4.90^{+0.55}_{-0.52}$ & $70^{+9}_{-7}$ & $1700^{+410}_{-390}$ & $5.8$ \\
 
 & ($6.393 - 6.401$) & ($4.21 - 5.64$) & ($60 - 80$) & ($1180 - 2250$) & $1.2$ \\
& & & & & \\
\hline
\end{tabular*}
}
{\small
{\it Chandra} HEG data, fitted with a power law plus Gaussian \fekalfa emission-line
model in the 2--7 keV band (see \S\ref{hegspec}).
The purpose of these fits is
primarily a {\it comparison} of the \fekalfa emission-line parameters
(see \figcontmodelsp, \figcontsnapshotsp, and \figcompshoulder for more complex
fits).
Absolute values of the line intensity are model-dependent but the
peak energy and width of the \fekalfa line core are insensitive to
the continuum and absorption model.
The last row gives measurements from the summed spectrum (with
$\sim 830$~ks exposure time), as derived in Yaqoob \& Padmanabhan (2004).
All fit parameters are quoted in the source rest
frame. Line widths are {\it intrinsic}, since they
have already been corrected for the instrumental
line-response function.
Statistical errors are for three interesting parameters
and correspond to the 68\% confidence
level ($\Delta C$-statistic $=3.51$), whilst parentheses show the
90\% confidence level ranges of the parameters ($\Delta C$-statistic $=6.25$).
$^{a}$  Gaussian line center energy.
$^{b}$  Emission-line intensity in units of $\rm 10^{-5} \ photons \ cm^{-2} \ s^{-1}$.
$^{c}$  Emission line equivalent width.
$^{d}$  Full width half maximum, rounded to $5 \ \rm km \ s^{-1}$.
$^{e}$  $F$ is the
estimated 2--10~keV observed flux in units of $10^{-11} \ \rm ergs\ cm^{-2}\ s^{-1}$
using the complex disk-line model in \S\ref{diskfitting}.
The power-law continuum was extrapolated to 10 keV.
$L$ is the estimated 2--10~keV source-frame luminosity
(using the 2--10 keV estimated flux), in units of $10^{43} \ \rm ergs\ s^{-1}$, assuming
$H_{0} = \rm 70 \ km \ s^{-1} \ Mpc^{-1}$, $\Omega_{\rm matter}=0.3$ and $\Omega_{\Lambda}=0.7$.
}
\end{table*}

At least part of the 
\fekalfa fluorescent 
emission line in type~I active galactic nuclei (AGNs)
is believed to originate in a
relativistic accretion disk around a black hole
(e.g. see reviews by Fabian \etal 2000; Reynolds \& Nowak 2003).
The dominant peak energy of the \fekalfa line 
at $\sim 6.4$~keV 
appears to be ubiquitous and this core of the line carries a
substantial fraction of the total line flux 
(e.g. Nandra \etal 1997; Sulentic \etal 1998; Lubi\'{n}ski \& Zdziarski 2001; 
Weaver, Gelbord, \& Yaqoob 2001; Yaqoob \etal 2002;
Perola \etal 2002; Reeves 2003; Yaqoob \& Padmanabhan 2004). 
It has been traditional to associate such narrow \fekalfa lines   
with an origin in distant matter, at least several
thousand gravitational radii from the putative black hole (e.g.
the optical broad-line region (BLR), the
putative obscuring torus, or the optical narrow-line region (NLR)).
However, Petrucci \etal (2002) recently reported a {\it variable},
narrow \fekalfa line in Mkn~841, supporting an accretion-disk origin
(see also Longinotti \etal 2004).
Moreover, rapidly variable, narrow Fe K line emission 
has been observed in the Seyfert~I galaxy NGC~7314 (Yaqoob \etal 2003a).
Thus, even narrow \fekalfa lines may have
a significant contribution from the accretion disk
(Lee \etal 2002; Yaqoob \etal 2003a; Longinotti \etal 2004; 
Turner, Kramer, \& Reeves 2004).

NGC~3783 is a fairly bright 
($F_{2-10 \rm \ keV} \sim 5-7.5 \times 10^{-11} \rm \ ergs \ s^{-1}$),
moderate luminosity ($L_{\rm 2-10 \ keV} \sim 1-1.5 \times 10^{43}
\rm \ ergs \ s^{-1}$), nearby ($z=0.00973$) Seyfert~1 galaxy which has
been well-studied in all wavebands. In the X-ray band, \src was the
target of the deepest observation with the 
\chandra high energy grating transmission spectrometer ({\it HETGS})
for any Seyfert galaxy, during a campaign in 2001, obtaining a
net exposure time of $\sim 830$~ks. The spectra from this campaign
represent the highest spectral resolution data with the best signal-to-noise
for any Seyfert galaxy available to date. The observing campaign was 
also designed to study variability and was broken up into five 
observations 
separated by various intervals ranging from days to months.
The soft X-ray part of the \hetg data has been well-studied 
by several research groups and has produced a wealth of new information
and insights into the photoionized outflow and its variability
(Kaspi \etal 2001, 2002, Netzer \etal 2003, Krongold \etal 2003). 
These results have
been supplemented by studies using \xmm (Behar \etal 2003; 
Reeves \etal 2004), 
and by studies in the UV band 
(e.g. Gabel \etal 2003a,b) 
On the other hand, the high-spectral resolution \hetg data 
for the \fekalfa line in NGC~3783 from the extended campaign in 2001 have
not yet been fully exploited. Kaspi \etal (2002) reported measurements
of the narrow core of the \fekalfa line and detection of a
Compton-scattering ``shoulder'' on the red-side of the core, indicating
an origin for the line core in cold, optically-thick matter, far from the
nucleus (beyond the optical BLR). Results of 
variability studies of the \fekalfa line from
within the extended observation campaign in 2001 have not yet been
reported. A detailed study of the \fekalfa line in \src using \xmm
was presented by Reeves \etal (2004) who confirmed 
the \chandra measurements of the narrow core, and in addition reported
that a broad, relativistic disk line was not required by the data
(a similar conclusion was reached by Kaspi \etal (2002) 
for the \chandra \hetg data).

In this paper we specifically study the \fekalfa line
in detail in
NGC~3783 using the \chandra \hetg data from the extended campaign in 2001. 
We study the time-averaged spectrum, as well as the individual spectra
from the five observations from the campaign in order to address 
line variability. We find that a Compton shoulder model is
not a unique description of the complexity in the \fekalfa line.
A relativistic disk line model provides as good a fit
as the Compton shoulder model.
In \S\ref{data} we describe the data and observations. In \S\ref{hegspec}
we discuss the results of spectral fitting to derive
parameters for the \fekalfa line core, from the time-averaged spectrum
as well as from the separate observations. In \S\ref{heabs} we confirm
the detection of an Fe He-like absorption feature.
In \S\ref{shoulderfit}
and \S\ref{diskfitting} 
we describe the results of fitting a Compton shoulder model and
relativistic disk-line model respectively, to the excess flux at the
base of the core
of the \fekalfa line. Finally, we present our conclusions in \S\ref{conc}.

\section{OBSERVATIONS AND DATA}
\label{data}
NGC~3783 was monitored over a period of $\sim 125$ days,
starting 2001 February 24, with the \chandra \hetg
during five snapshots. We will refer to these snapshots as
observations (1) to (5). 
\hetg consists of two grating assemblies,
a High-Energy Grating (HEG) and a Medium-Energy Grating (MEG),
and it is the HEG that achieves the highest spectral resolution.
The MEG has only half of the spectral resolution
of the HEG and less effective area in the Fe-K band, so our study will
focus on the HEG data.
The \chandra data 
were reduced and HEG spectra made exactly as described in Yaqoob \etal (2003b).
We used only the first orders of the grating data (combining
the positive and negative arms). 
The mean HEG count rates ranged from $0.36-0.50$ ct/s.
The exposure times for observations (1) to (5) were
in the range 165--169~ks. The count rate and exposure time
for the total time-averaged HEG spectrum were 0.36~ct/s and
832~ks respectively.
Background was not subtracted since it is negligible over the
energy range of interest (e.g. see Yaqoob \etal 2003a).
Note that 
the systematic uncertainty in the HEG
wavelength scale 
is $\sim 433 \ \rm km \ s^{-1}$ ($\sim 11$~eV) at 6.4 keV
\footnote{http://space.mit.edu/CXC/calib/hetgcal.html}. 

\begin{figure*}[bth]
\vspace{10pt}
\centerline{\psfig{file=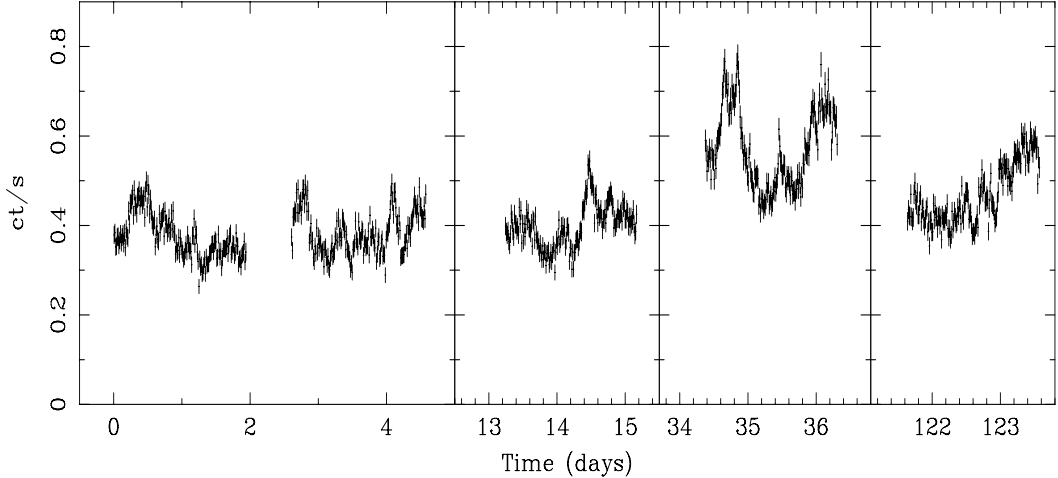,width=5.5in,height=2.5in}}
\vspace{-8mm}
\caption{\footnotesize Lightcurves from the \chandra high-energy grating
({\it HETGS}) monitoring campaign of NGC 3783.
In this paper, the five snapshots are referred to as observations (1)--(5)
The first panel shows the first {\it two} snapshots, and the remaining
three panels show one snapshot in each panel.
Shown are the 2--7~keV count rates combined from MEG and HEG data,
utilizing both the $-1$ and $+1$ orders of the gratings, binned at 1024 s.
The zero reference time in this composite lightcurve is
UT 2001 February 24 19:06:15. The scale on the time axis is such
that equal lengths correspond to equal time intervals.
}
\end{figure*}

\figlc shows the 2--7~keV lightcurves for the five 
observations, showing the count rates summed over the $-1$
and $+1$ orders of both the MEG and the HEG, binned at 1024 s.
It can be seen that the observations, each of duration $\sim 2$ days,
were designed to 
probe timescales of $\sim 1$ day and less, $\sim 1,2,3$, and
4 weeks, 3 months, and 4 months.
The 2--7~keV continuum variability over the first $\sim 2$ weeks of
the campaign was confined to the
flux remaining within $\sim \pm 25\%$ of the
mean during that period, but in observation (4),
about a month into the campaign, the overall flux was higher
than the mean in the first two weeks of the campaign by $\sim 50\%$ 
(see \figlcp).
However, the variability about the new mean flux during observation (4)
was still $\sim \pm 30\%$ relative
to the mean in observation(4). In the final observation the 2--7~keV flux
began back at its level at the beginning of the campaign, but 
by the end of the $\sim 2$ day observation it had risen by $\sim 50\%$,
compared to the level at the beginning of final observation of the campaign.
The excess variance above the
expectation for Poisson noise
(e.g. Turner \etal 1999), calculated over the entire $\sim 4$ month monitoring
period from the lightcurve in \figlc 
is $(4.52 \pm 0.28) \times 10^{-2}$.
 
\section{SPECTRAL FITTING RESULTS}
\label{hegspec}

We used XSPEC v11.2 (Arnaud 1996) for spectral fitting.
Since we were interested in utilizing the highest possible
spectral resolution available, we used spectra binned
at $0.0025\AA$, and this amply over-samples the HEG resolution
($0.012\AA$ FWHM).
The $C$-statistic was used for minimization.
By definition, calculation of the
$C$-statistic requires only knowledge of the number of counts
in a bin, but for spectral plots, the error bars shown
correspond to asymmetric errors calculated using the approximations
of Gehrels (1986).
All model parameters will be
referred to the source frame, unless otherwise noted.
Note that since all models were fitted
by first folding through the instrument response before comparing
with the data, the derived model parameters {\it do not}
need to be corrected for instrumental response.

\subsection{Simple Continuum Model}
\label{hegsimplespec}

Our method is first to fit a simple empirical model
to the continuum and extract parameters for the core of the \fekalfa 
emission line to establish whether there is any intra-observation
variability of the line parameters. 
Since we are {\it comparing} line parameters, and the line core
is narrow (see YP04), this is good enough to guide subsequent,
more detailed analysis. Accordingly, we will then describe more physical
(and more complex) models for the continuum and absorption and show the
extent of the sensitivity of the
absolute parameters of the \fekalfa emission-line core to
details of these models.

Thus, we fitted
a simple power law plus Gaussian
emission-line model over the 2--7~keV band for each of the
five spectra.
The Gaussian component had three free
parameters (line center energy, width, and intensity)
so the model had
five free parameters in total, including the continuum slope
and normalization. 

Although the \fekalfa line consists of two components
($K\alpha_{1}$ and $K\alpha_{2}$, separated by 13~eV), we
modeled it as a single Gaussian, since it was shown in Yaqoob \etal (2001),
that with the spectral resolution of the HEG, there
is a negligible impact on the measured line width. 
Some broadening may also result from the presence of
line emission from more than one ionization state of Fe. However, we
will bear this in mind when interpreting 
the measured FWHM velocities.
Also, the use of a single Gaussian (without any attempt to
model the underlying broad \fekalfa emission) has a negligible
impact on the measured center energy of the core
(see \S\ref{diskfitting} and Yaqoob \etal 2001). 
In any case, we will compare the parameters obtained from
the simple model with those obtained from more complex models
(see \S\ref{contabsdepend}, \S\ref{shoulderfit}, and \S\ref{diskfitting}).

The best-fitting emission-line parameters for each spectrum
are shown in \tablefits (as well as extrapolated 2--10~keV
fluxes and luminosities obtained using more complex models for the
line and continuum, as described in \S\ref{diskfitting}). 
In order that the results can used for future statistical analyses,
statistical errors  
corresponding to 68\% confidence for three interesting parameters
($\Delta C = 3.506$) are given. In addition,
a more conservative measure,
the 90\% confidence, three-parameter range 
($\Delta C = 6.251$) for each line parameter is also given 
in \tablefitsp.

\tablefits shows that the peak energy of the \fekalfa emission-line
core is tightly constrained around 6.40~keV in all five observations,
and shows no evidence for variability even at the 68\% confidence
level. The line intensity has 68\% confidence statistical errors
of the order of $\sim 30\%$ and also shows no sign of variability.
The equivalent width (EW) however, ranging from $\sim 50-90$~eV, does
show variability at the 68\% confidence level, but not at the
90\% confidence level. However, any EW variability in this case
is simply due to the continuum level changing whilst the line
intensity remains steady. The 2--10~keV flux and luminosity show
only $\sim 50-60\%$ variability over the entire monitoring campaign.
The most interesting result from \tablefits is that in the first
two observations the \fekalfa line core appears to be resolved 
at the 90\% confidence level, but in the latter three observations
it appears to be unresolved. The 90\% upper limit on the FWHM in
the last three observations is not more than $3000 \rm \ km \ s^{-1}$,
but for the first two observations it is $4000 \rm \ km \ s^{-1}$
or more. For comparison, the FWHM resolution of the HEG at 6.4~keV
is $\sim 1860  \rm \ km \ s^{-1}$. 

We report the results of investigating this apparent variability 
of the width of the
\fekalfa line core further in \S\ref{linevar}, but first we 
report the results of investigating
the dependence of the \fekalfa core line parameters on the details of modeling
the continuum and absorption.

\begin{figure*}[bth]
\vspace{10pt}
\centerline{\psfig{file=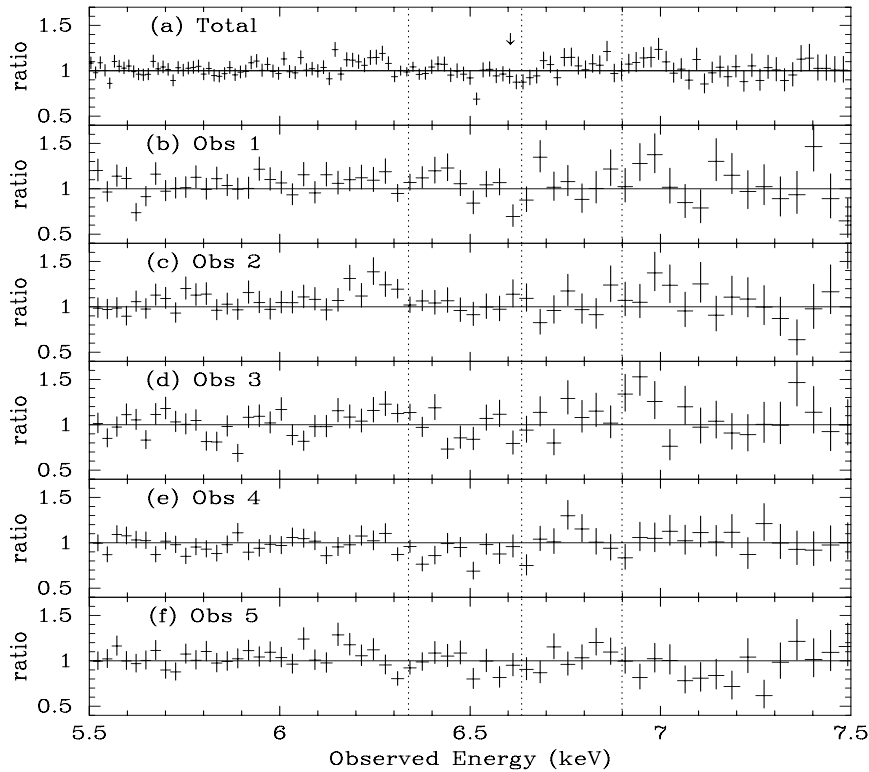,width=4.5in,height=4.0in}}
\vspace{-8mm}
\caption{\footnotesize Ratios of data/model for the \chandra HEG data for NGC 3783.
The model consists of a complex continuum plus warm absorber
and simple Gaussian \fekalfa emission line, fitted as described in
\S\ref{contabsdepend}, to the total
$\sim 830$~ks HEG spectrum (i.e. all snapshots combined).
The ratio of the total HEG spectrum, binned at $0.005\AA$,
to this model is shown in (a).
The ratios of HEG spectra (for the individual
snapshots binned at $0.01\AA$) to exactly the same model
(with all parameters fixed at the values obtained from fitting
the total spectrum, apart from an overall renormalization) are
shown in (b)--(f). Note that the bin size for the
individual snapshot spectra in (b)--(f)
is comparable to the HEG FWHM spectral resolution
of $\sim 0.012\AA$. It can be seen that there is very little
evidence for spectral variability, aside from line-like residuals
between $\sim 6-7$~keV in some of the ratio plots. Vertical dashed lines
indicate energies of 6.400~keV, 6.700~keV, and 6.966~keV in the NGC 3783
rest frame,
corresponding to the energies of the Fe~{\sc i}~$K\alpha$, \fexxv \resonetwo
(resonance), and \feklya transitions. The arrow in (a) indicates the
position in energy of the Fe-K absorption feature found in the
\xmm data by Reeves \etal (2004).
}
\end{figure*}

\subsection{Dependence of \fekalfa Line Parameters on Continuum and 
Absorption Modeling}
\label{contabsdepend}

Here we use the $\sim 830$~ks MEG and HEG spectra (i.e.
summed over the five observations) to investigate in detail
the effect of using an over-simplified empirical
continuum (as described above) on the derived \fekalfa 
core line parameters.
It is well known that there is a complex photoionized absorber
in NGC~3783, with at least two components. The absorber has
been modeled in great detail by several research groups,
using \chandra and \xmm data (e.g. Kaspi \etal 2002; Netzer
\etal 2003; Krongold \etal 2003; Behar \etal 2003).
Since highly ionized components of the absorber can affect
the data even in the Fe-K band, it is conceivable that the
warm absorber can affect interpretation of the 
broad part of the \fekalfa emission line.
In fact, the apparent lack of a {\it broad} \fekalfa emission line
in recent \xmm data was attributed to the complex absorber not
modeled adequately in previous analyses (Reeves \etal 2004).
Reeves \etal (2004) also reported an absorption line due
to He-like or H-like Fe in the same \xmm data.
We expect the effect of the photoionized absorber on the 
narrow \fekalfa emission-line core to be unimportant, but since
we will also be searching for spectral variability over a 
broader energy range than just the line core, and
revisiting the question of the presence of a broad \fekalfa line, 
we constructed a model of
the photoionized absorber.

We compiled a spectral energy distribution (SED) from the
literature to use as an input to the photoionization code 
XSTAR\footnote{{\tt http://heasarc.gsfc.nasa.gov/docs/software/xstar/xstar.html}},which was then used to construct grids of models
for spectral fitting. Full details and
results for the complex absorber modeling are given in 
McKernan, Yaqoob, \& Reynolds 
(2005, MN, submitted). For the present purpose of
\fekalfa line modeling, we did not attempt
to model the individual soft X-ray absorption-lines
and features at the level of
detail accomplished in the dedicated studies mentioned above.
We found that a two-component absorber combined with a 
broken power-law continuum was sufficient to model
the broadband MEG and HEG data. The model was derived by first
fitting the MEG 0.5--7~keV data, using an additional Gaussian
to model the \fekalfa emission-line core, whose parameters
were initially frozen at the values derived using the simple
power-law continuum model (\S\ref{hegsimplespec}).
Then this model was fitted to the 0.8--9~keV HEG data
with all parameters except the hard X-ray slope frozen.
Then, restricting the energy range to 2--7~keV
(in order to directly compare derived emission-line parameters
with the empirical continuum fits in \S\ref{hegsimplespec})
the Gaussian emission-line parameters (peak energy, width,
and intensity) were allowed to float and the best-fit found.
The photoionized absorber components had column densities of
$4.7 \times 10^{21} \ \rm cm^{-2}$ and
$3.5 \times 10^{22} \ \rm cm^{-2}$, with corresponding
ionization parameters $\log{\xi} = 0.74$ and 2.21 respectively
(where $\xi \equiv L_{\rm ion}/n_{e} r^{2}$,  
$L_{\rm ion}$ being the 1--1000 Rydberg ionizing luminosity,
$n_{e}$ the electron density, and $r$ the source to absorber
distance).
The soft X-ray modeling is discussed in detail in McKernan
\etal (2005, MN, submitted), as well as in the several studies mentioned
above so we do not discuss it any further. \figsixspecrat(a) shows
the ratio of the summed HEG data to this model in the 2--7~keV
band, with the data binned at $0.005\AA$. 
The other five panels
in \figsixspecrat (b)--(f)
show the ratio of the data, from each of the five
snapshots, {\it to this same model, with only the overall
normalization adjusted to obtain the best-fit}.
Aside from   
line-like features between $\sim 6-7$~keV,
the model gives a strikingly good fit to the spectra from
all five of the separate observation data sets.
We will address the
possible origin of these residuals in more detail in 
\S\ref{diskfitting}.
The apparently
variable line-like features at $\sim 7$~keV are likely to be due to 
\feklya emission
(which, blended with Fe~K$\beta$, was detected
in \xmm data by Reeves \etal 2004;
also see Kaspi \etal 2002).

\begin{figure*}[htb]
\vspace{10pt}
\centerline{\psfig{file=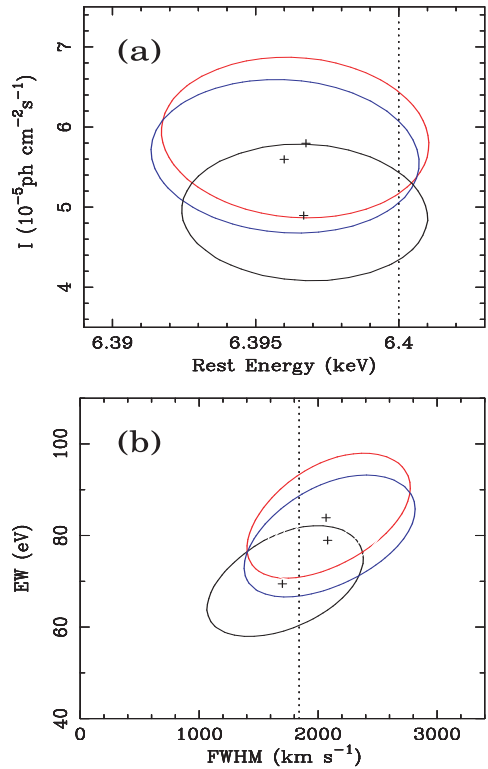,width=2.5in,height=4.0in}}
\vspace{-8mm}
\caption{\footnotesize
Joint, two-parameter, 99\% confidence contours for the \fekalfa emission-line
core from the total, $\sim 830$~ks, \chandra HEG spectrum of
NGC~3783, illustrating the effects of different levels
of approximation in modeling the continuum. Black contours correspond
to a simple power-law continuum (fitted in the 2--7~keV
band), red contours correspond to a model including
a photoionized absorber, as described in \S\ref{contabsdepend}, and blue
contours include a Compton-reflection continuum in addition
to the photoionized absorber, also described in \S\ref{contabsdepend}.
The emission-line is modeled by a simple Gaussian and (a) shows the line 
intensity versus center energy, and (b) shows the line equivalent width (EW) 
versus FWHM. All quantities are in the rest frame of NGC~3783.
It can be seen that the center energy of the line is not
sensitive to details of modeling the continuum, but there is
some model dependency on the line intensity, EW, and FWHM.
However, the differences are not statistically significant
at the 99\% confidence level. The dotted line in (a) corresponds
to the rest-energy of Fe~{\sc i}~$K\alpha$ (6.400~keV) and the
dotted line in (b) corresponds to the HEG FWHM resolution at
the observed center energy of the \fekalfa line.
}
\end{figure*}

\figcontmodels shows joint, two-parameter, 99\% confidence
contours directly comparing the derived \fekalfa line results
for the simple and complex continuum models. Black contours
show the results using the simple, empirical power-law
continuum, and red contours show the results using the
warm absorber plus complex intrinsic continuum described above.
Specifically, \figcontmodels(a) shows the \fekalfa line intensity
versus line center energy contours, whilst \figcontmodels(b)
shows the EW versus FWHM contours. It can be seen that although
both the line intensity and EW are consistent at the
99\% confidence level from the two
different continua, the fits using a complex continuum
give values for the best-fitting line intensity and EW which are $\sim 20\%$
higher than those obtained using the simple power-law continuum only.
This is because the intrinsic flux from the line has to
be larger to compensate for the absorption in the complex
model. However, we do not in fact know whether the warm absorber
lies farther from the X-ray continuum source than the \fekalfa line
emitter so there is  an inherent uncertainty in the
\fekalfa line intrinsic intensity which will only be resolved
when we are able to constrain the geometry of the system better
with future missions (but see \S\ref{shoulderfit} and \S\ref{diskfitting}). 

The \chandra HEG data are not sensitive enough to
strongly constrain a 
Compton-reflection continuum.
However, we investigated the effect of
including a Compton-reflection continuum 
(in addition to the complex ionized absorber
model described above) since it is conceivable
that it could still affect the derived line parameters as it
has a complex shape in the 6--7~keV region due to the Fe-K edge.
Moreover, if the \fekalfa emission line is formed in optically-thick
matter (for example reflection from the inner surface
of a putative obscuring torus), a Compton-reflection continuum
is {\it expected}. We used the {\sc pexrav} model in XSPEC
(see Magdziarz and Zdziarski 1995). Although this
model is based on a disk geometry, it is adequate for
our purpose since the \chandra data are not sensitive
to the differences in geometry. Again, since the data cannot
constrain this model well, the parameters of the reflection model
were fixed at nominal values (except for the photon index and
normalization of the intrinsic power-law continuum).
Specifically, we fixed the Fe abundance at the
default solar value, the disk inclination angle at $30^{\circ}$,
and the effective solid angle of the reflector at
$2\pi$ (i.e. the ``reflection factor'', $R=1$). 
However, in \S\ref{shoulderfit} and \S\ref{diskfitting} we
will investigate the effect of allowing $R$ to be free.
The e-folding
cutoff energy of the power law was fixed at 300~keV, well
outside the range of the data.

\figcontmodels(a) shows the joint, two-parameter,
99\% confidence contour (blue) of the \fekalfa line intensity versus line
center energy, and \figcontmodels(b) shows the
two-parameter, 99\% confidence contour of the line EW versus FWHM.
These
contours are directly compared with the corresponding contours
derived with the Compton-reflection continuum omitted (red), and with
only a simple power-law (black). It can be seen that the effect
of omitting the Compton-reflection continuum on the derived
parameters of the core of the \fekalfa emission line is negligible,
even at the 99\% confidence level.
The effect of the ionized absorber on the line parameters is
far more important.

\subsection{Variability of the \fekalfa Line Core}
\label{linevar}
\figcontsnapshots(a) shows the joint, two-parameter,
99\% confidence contours of the \fekalfa line intensity versus
center energy obtained from the time-averaged HEG spectrum
(black) compared with those obtained from the five
individual observations (or ``snapshots'').
This confirms the result from the spectral fitting (see \tablefitsp),
that at 99\% confidence, the intensity and peak energy of
the \fekalfa line core do not vary.

The five snapshots span a range of timescales from
$<2$ days to $\sim 125$ days. 
However, long-term intensive monitoring of \src with \rxtep,
some of which was
simultaneous with the present \chandra observations, shows
that the continuum varies significantly on all timescales in this
range (Markowitz \& Edelson 2004). Therefore all we can say is
that the actual response time of the \fekalfa line core must be greater than
$\sim 2$ days because if the continuum varies more rapidly than the
response time, the \fekalfa line core intensity measured will correspond
to some average continuum level. If we take $t>169$~ks as the
response time for the \fekalfa line core to continuum variations,
we get $r>5 \times 10^{15}$ cm for the location of the line-emitter
relative to the continuum source
(the recombination timescale likely does not play a role
here since the center energy of the \fekalfa line
core indicates neutral Fe). For
a central black hole mass of $2.9 \times 10^{7} \ M_{\odot}$
(Peterson \etal 2004), this corresponds to $r>1000r_{g}$, where
$r_{g} \equiv GM/c^{2}$ is the gravitational radius.
However, the FWHM of the \fekalfa line core ($\sim 1700 \rm \ km
\ s^{-1}$ from the total combined spectrum) places its origin
much further from the central black hole. 
The actual width already indicates that it might coincide with
the outer BLR/inner NLR. Note that in the present paper,
all quoted line widths derived from
spectral fitting are already corrected for the instrumental
line-spread function. 
Assuming a virial
relation and an r.m.s. velocity dispersion of $\sqrt{3} v_{\rm FWHM}
/2$ (e.g. Netzer 1990), 
places the line emitter at 0.06 pc, or $\sim 70$ light days
from the central continuum source
(using the above mass). If some of the line width is not
kinematic in origin (e.g. if there are multiple lines from different
ionization states of Fe), then this distance is even greater.
This is consistent with the lack of response of the line intensity
to continuum variability. However, the lack of variability of
the narrow \fekalfa line core {\it is not necessarily an
expected result}, because variability in the narrow \fekalfa line
core has been observed in Mkn~841 (Petrucci \etal 2002),
independent of the continuum (and that result is still not
fully understood, but see Longinotti \etal 2004).

In \S\ref{hegsimplespec} (\tablefitsp) we showed that observations
(1) and (2) appeared to have a somewhat broader \fekalfa line core
than observations (3)--(5). To investigate further, we combined
the five HEG spectra into two groups, one spectrum from observations
(1) and (2), and the other spectrum from observations (3)--(5).
\figcontsnapshots(b) shows the joint, two-parameter,
68\%, 90\%, and 99\% confidence contours of the \fekalfa line
EW versus the FWHM, for these two grouped spectra, compared with
the 99\% contour obtained from the mean, time-averaged spectrum.
Indeed, the 68\% confidence contours from observations (1) and (2)
do not overlap with those from observations (3)--(5), the latter
showing a narrower line width than the former, with the narrower line
having a somewhat smaller flux than the broader line.
Thus, at 68\% confidence, the \fekalfa line core appears to
be broader and more intense in observations (1) and (2) than in
observations (3)--(5).
However, the 90\% and 99\% confidence contours {\it do} overlap
so we cannot say that the line width varied at 90\% confidence or
greater. On the other hand, the line is resolved by the
HEG at higher than 99\% confidence in observations (1) and (2)
but is not resolved at 99\% confidence in observations (3)--(5).
Future instrumentation will have to address the question of
variability in the line width.

\begin{figure*}[htb]
\vspace{-2mm}
\centerline{\psfig{file=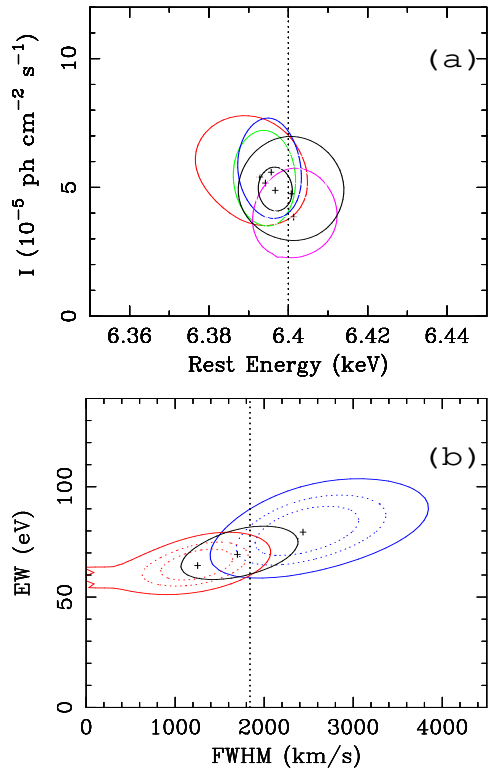,width=2.5in,height=4.0in}}
\vspace{-8mm}
\caption{\footnotesize
(a) Joint, two-parameter, 99\% confidence contours
of line intensity versus center energy (both in the AGN rest frame)
for the \fekalfa emission-line
core (modeled with a simple Gaussian) from \chandra HEG spectra of
NGC~3783. The continuum was modeled by a simple power-law,
fitted over the 2--7~keV energy range. The small, black contour was obtained
from the total, $\sim 830$~ks, HEG spectrum. Remaining contours
were obtained from the individual snapshots as follows: black (large),
observation (1); red, observation (2); green, observation (3); blue,
observation (4); magenta, observation (5). See \S\ref{hegspec} for details.
It can be seen that there is no discernible
variability at the 99\% confidence level. The dotted line is at
6.400~keV in the rest frame  of NGC~3783.
(b) Joint, two-parameter, confidence contours
of line equivalent width (EW) versus line FWHM (both in the AGN rest frame)
for the \fekalfa emission-line
core in the NGC~3783 HEG data, using the same model as in (a).
Black contour shows the 99\% confidence region
for the total, $\sim 830$~ks, HEG spectrum. Remaining contours
were obtained by combining the snapshots from observations (1) and (2)
(blue contour), and observations (3), (4), and (5) (red contour).
Dotted lines show the 68\%, and 90\% confidence regions and solid
lines show the 99\% confidence contours.
These combinations were formed after examining the contours
of all five observations separately and grouping them according
to similar widths.
It can be seen that there is marginal evidence for a change
in the effective width of the line, which appears to be
resolved in observations (3), (4) and (5), but not in observations
(1) and (2) (the dotted line corresponds to the HEG FWHM resolution
at the observed center energy of the line).
}
\end{figure*}

\section{He-like Fe Absorption Feature}
\label{heabs}

Kaspi \etal (2002) reported He-like Fe absorption in 
the \chandra \hetg data.
Reeves \etal (2004) detected the absorption line from
an \xmm observation of NGC~3783 (in December 2001), centered
at $6.67 \pm 0.04$~keV (in the source 
rest frame), with an EW of $17 \pm 5$~eV. 
The line was attributed to \fexxv \resonetwo
resonance absorption and it was not resolved by the CCDs. 
Evidence for this absorption feature can be seen in \figsixspecrat
(an apparent absorption dip is indicated by an arrow).
We modeled the feature by adding an inverted Gaussian
to the two-component warm absorber plus broken power-law model
described in \S\ref{hegspec} (including the Gaussian emission line).
All parameters
except for the hard power-law slope, the overall continuum normalization,
and the three absorption-line parameters  (center energy, intrinsic 
width) were frozen at their best-fitting values. 
The $C$-statistic decreased by 9.2 upon the addition of the
absorption line, which corresponds to a detection significance of
only 99\% for three additional free parameters.
For the line center energy, EW, and intrinsic width, we measured
$6.62^{+0.07}_{-0.06}$~keV, $12^{+8}_{-7}$~eV, and
$6405^{+5020}_{-2670} \rm \ km \ s^{-1}$ FWHM respectively
(errors are one-parameter, 90\% confidence, to facilitate direct
comparison with the \xmm results of Reeves \etal 2004).
Thus, the \chandra \hetg and \xmm measurements
are consistent with each other. 

\begin{figure}[tbh]
\vspace{10pt}
\centerline{\psfig{file=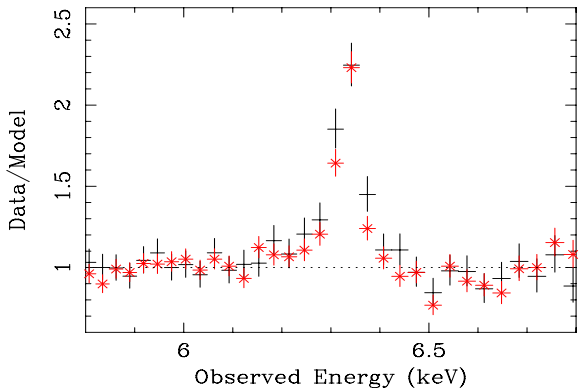,width=3.0in,height=2.0in}}
\caption{\footnotesize
Ratios of HEG data for NGC~3783
to a simple power-law continuum model, fitted in the 5--7~keV
band, omitting the 6.0--6.5~keV data.
Black data points correspond
to the summed spectra from observations (1) and (2), and the
red data points correspond to the summed spectra from
observations (3), (4), and (5). These two groupings show similar
line widths within each group, from examination of the data from
the individual observations (see \figcontsnapshots and \S\ref{hegspec}).
This direct comparison shows how the \fekalfa line core is
slightly broader in observations (1) and (2) (black) than in observations
(3), (4), and (5) (see \figcontsnapshotsp).
The continuum was fitted to the two data sets simultaneously,
with the photon index tied together. Although the continuum
is over-simplified, the purpose here is simply to compare the
emission-line core for the two data sets.
}
\end{figure}

\section{\fekalfa Line Compton Shoulder}
\label{shoulderfit}

Using the same NGC~3783 \chandra HEG data as described in the present paper,
Kaspi \etal (2002) noted the detection of a so-called ``Compton shoulder'' on
the red side of the peak of the \fekalfa emission line. This structure 
would be due to the core 
\fekalfa emission-line photons Compton-scattering on electrons
in the medium in which the line is formed, before escaping the medium
and reaching the observer. If the material in which the line is formed
is not too hot (i.e. $kT \ll 6.4$~keV), the Compton scattering 
results in the line photons losing energy so that the scattered emission
appears redward of the peak of the line. The relative magnitude of the
Compton shoulder relative to the peak line flux increases with
optical depth of the medium, until the medium becomes optically-thick
(see, for example, Matt 2002, for a theoretical description
and Monte Carlo calculations). The peak of the once-scattered photon
energy distribution is expected at two Compton wavelengths longer than
the wavelength of the unscattered line photons. For a line at 6.4~keV,
this is $\sim 6.25$~keV in the rest frame of NGC~3783, or 6.19~keV
in the observed frame. The excess flux to the red side of the \fekalfa line
core can be seen in \figsixspecrat(a) and \figtwofekrat
and does in fact appear to extend down to the appropriate energy. 

We applied a model of the Compton shoulder 
to the NGC~3783 HEG data. 
For the line scattering we used the XSPEC model of Watanabe \etal (2003).
The intrinsic width of the emission line in this model is much
less than the instrument resolution and includes
no kinematic information so is rather unphysical. During preliminary fitting
we found that the model always
left excess flux blueward of the \fekalfa line core.
Therefore we convolved the line
profile with a Gaussian whose width was a free parameter,
in order to mimic Doppler broadening of the intrinsic line profile.
We used the broken power-law plus two-component warm absorber, 
with a Compton-reflection
continuum as the baseline model, as described in \S\ref{contabsdepend}.
The parameters of
the warm absorber and Compton reflection
models were fixed at the values
described in \S\ref{contabsdepend}.
The fitting was performed between 2--9~keV
(i.e. the high-energy end of the fitted range was
now extended from 7~keV up to 9~keV).
Only the hard X-ray slope, the overall continuum normalization, and
the column density of the Compton-scattering medium were free parameters.
Since the \fekalfa line peak energy indicates 
low ionization states of Fe and is consistent with Fe~{\sc i}, 
the temperature of the scattering medium must be low
and we assumed the medium is neutral (``cold''). 
The data, best-fitting model, 
and data/model ratio are shown in \figcompshoulder (a).
Note that the He-like Fe absorption line discussed in
\S\ref{heabs} has deliberately not been modeled in 
\figcompshoulder (a) in order to clearly show its presence
and magnitude, especially in the data/model ratios, and
to facilitate comparison with the \xmm data. 
We obtained a FWHM of $1500^{+460}_{-340} \ \rm km \ s^{-1}$
for the width of the Gaussian used to convolve the 
Compton-scattered line profile. The 
column density we derived was $7.5_{-0.6}^{+2.7} \times 10^{23} \ \rm cm^{-2}$
(90\%, one-parameter errors).
This corresponds to a Thomson depth of $\sim 0.60$, which is
not quite Compton-thick.
The Thomson depth of the scattering medium must be less than
$\sim \sqrt{2}$, or else the mean number of scatterings would be $\sim 2$ and
the peak energy of the scattered photons would be much lower than
measured (the scattered line photons would have a distribution with
significant flux down to 
four Compton wavelengths below the zeroth-order line peak).
A Thomson depth of $\sqrt{2}$ 
corresponds to $\sim 1.7 \times 10^{24} \ \rm cm^{-2}$. 

Ideally, we would self-consistently model the Compton-reflection
continuum and the Compton-scattered line since both are
produced by the same physical process. When we allowed the
relative normalization of the Compton reflection continuum, $R$,
to be a free parameter, we obtained $R=1.0^{+0.5}_{-0.3}$
(justifying $R=1$ in the fits thus far). However, the
Compton-reflection continuum model is calculated for an
optically-thick disk, but the emission line may be produced in
transmission
since we measured a small Thomson depth from the Compton shoulder.
Better data are required to justify more
sophisticated modeling.

\begin{figure*}[bth]
\vspace{-5mm}
\centerline{\psfig{file=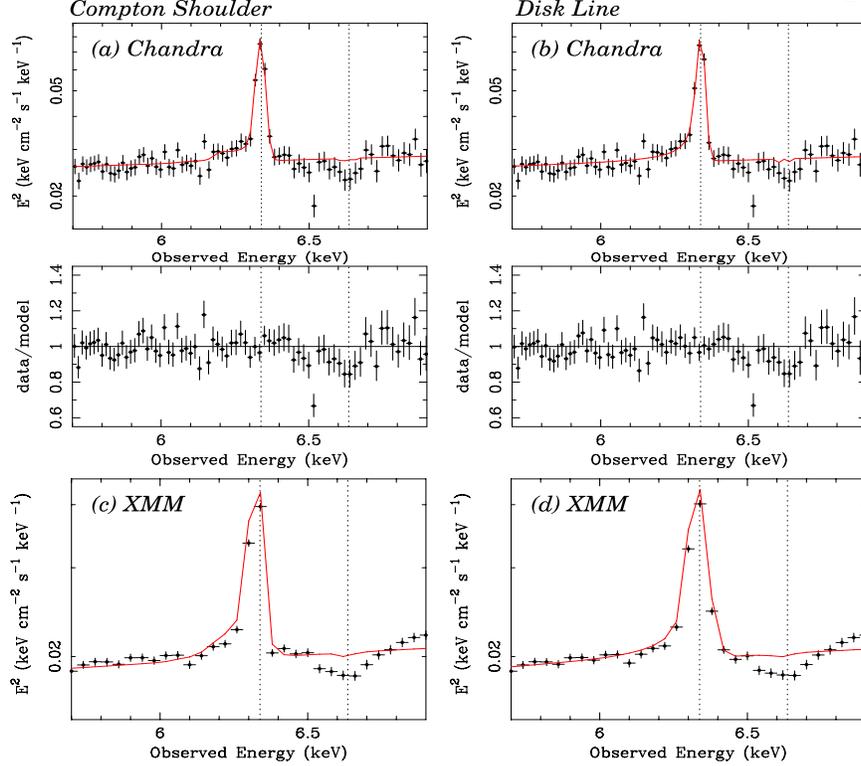,width=4.5in,height=4.0in}}
\vspace{-10mm}
\caption{\footnotesize(a) Modeling of the Compton shoulder of 
the \fekalfa line in the
total, $\sim 830$~ks, \chandra HEG spectrum of NGC~3783
(the data are binned at $0.005\AA$ for the plot).
The model consists of a power law, photoionized absorber, a
Compton-reflection continuum, and a narrow \fekalfa line which
suffers Compton scattering in cold, solar-abundance
material, observed in transmission.
The line profile was convolved with a Gaussian, whose
best-fitting width was $1500^{+460}_{-340} \rm \ km \ s^{-1}$.
See \S\ref{shoulderfit} for details
of the fitting procedure and results.
(b) The  \chandra HEG spectrum, same as in (a), fitted with
an emission line profile from a relativistic accretion disk
rotating about a Schwarzschild black hole, plus a Gaussian
to model the \fekalfa line core. See \S\ref{diskfitting} for
model-fitting details and results.
Note also that the dip at $\sim 6.51$~keV (observed frame)
is not real: it is narrower than the instrument resolution
($0.012\AA$ FWHM) and is not detected in {\it both} the $-1$ and
$+1$ orders of the grating.
(c) The \chandra
Compton-shoulder model in (a) fitted to non-contemporaneous \xmm
EPIC pn CCD data
for NGC~3783, with only the overall normalization and power-law
slope free. (d) The \chandra disk line plus Gaussian model in (b)
fitted to non-contemporaneous \xmm pn data
for NGC~3783, with only the overall normalization and power-law
slope free. See Reeves \etal (2004) for details
of the \xmm data.
We note that all of the line profiles shown here are
{\it not} unfolded: the profiles correspond to the ratio of
counts to predicted model counts, multiplied by the best-fitting
model.
Note that the absorption line at $\sim 6.7$~keV
was deliberately {\it not} modeled for any of the four
cases shown here, in order that the
reader can readily compare the absorption line
in the \chandra and \xmm data. See \S\ref{heabs}
or Reeves \etal (2004) for
detailed fitting of the absorption line in the \chandra
or \xmm data respectively.
}
\end{figure*}

\section{Relativistic Disk Line Model}
\label{diskfitting}

We investigated whether a relativistically broadened \fekalfa line
(in addition to a narrow Gaussian  component) could also account for
the \fekalfa line profile in NGC~3783. We fitted the 2--9~keV HEG
data using a 
broken power-law plus two-component warm absorber, with a Compton-reflection
continuum as the baseline model, as described in \S\ref{contabsdepend}.
Initially, the Compton-reflection parameter, $R$, was fixed
at unity, which corresponds to the expected steady-state normalization
of the reflected continuum from a neutral (``cold'') Compton-thick
disk subtending a solid angle of $2\pi$ at the X-ray source.
Preliminary spectral fitting with $R$ free showed that the best-fitting
value is $R \sim 1$ in any case (as in \S\ref{shoulderfit}), 
and we will give statistical
errors when $R$ was a free parameter below.
In addition to the above model components we
included a Gaussian (to model the line core), and an emission line from
a relativistic disk around a Schwarzschild black hole (e.g. see Fabian \etal 1989).
The energy, intrinsic width, and intensity 
of the Gaussian were free parameters.
The inner radius of line emission from the disk was initially fixed at $6r_{g}$
($r_{g} \equiv GM/c^{2}$) and the outer radius was fixed at $1000r_{g}$.
The line radial emissivity was assumed to be a power law
(line emissivity proportional to $r^{-q}$), with the index, $q$, free.
In the disk rest-frame the emission-line energy was fixed
at 6.4~keV. The disk inclination angle and 
overall line intensity were also free
parameters. Thus, there were a total of seven free parameters
for this model, including the continuum slope and normalization.

The data, best-fitting model, and data/model
ratio are shown in \figcompshoulderp(b). It can be seen
that the fit is as good as that for the Compton-scattering model,
shown in \figcompshoulderp(a).
Again, we deliberately did not model the He-like Fe absorption
feature discussed in \S\ref{heabs} in order to clearly
show its effect and to facilitate comparisons between
\chandra and \xmm data.
In the 5.7--6.9~keV energy band shown in
the figure, the $C$-statistic for the
disk-line model is less than that 
for the Compton shoulder model (\S\ref{shoulderfit}) 
by only 1.6
for an additional three free parameters.
Thus, although the disk-line model is not statistically
preferred over the Compton shoulder model, it cannot be ruled out.
The center energy of the Gaussian was 6.399~keV, with 
statistical errors similar to those obtained from simpler
continuum models (\tablefitsp). Since the peak energy was well-determined,
this parameter was fixed when deriving the statistical errors
for the other parameters in the model
(in this section, all errors
are 90\% confidence, one-parameter, or $\Delta C=2.706$). 
The Gaussian line FWHM obtained
was $1460^{+470}_{-680} \ \rm km \ s^{-1}$.  
The Gaussian line width was then fixed at the best-fitting value
in order to derive statistical errors on the remaining model
parameters, in order to avoid the spectral
fitting becoming unstable.

We obtained EWs of $56^{+14}_{-11}$~eV and $39^{+18}_{-16}$~eV for the
Gaussian and disk emission-line components respectively
(all line measurements are in the source frame).
The radial emissivity index measured for the disk-line emission
was $q = 1.90^{+0.63}_{-0.57}$, and the inclination angle 
was constrained with an upper limit only, of $11^{\circ}$. When
we allowed the relative normalization of the Compton-reflection
continuum, $R$, to be a free parameter we obtained $R=1.0^{+0.4}_{-0.3}$.
We recall that $R=1$ corresponds to the case of time-steady 
illumination of a thick disk subtending $2\pi$ solid angle at the
X-ray source.
We note that De Rosa \etal (2002) obtained $R=0.71^{+0.20}_{-0.28}$
from non-contemporaneous \bsax data
for NGC~3783, and Markowitz, Edelson, \& Vaughan (2003) obtained $R=0.62\pm 0.12$
from \rxte data
which overlapped with the first three \chandra
observations reported in the present paper, consistent, within the
errors, with the \chandra value.
However, Markowitz \etal (2003) also reported $R = 0.47^{+0.05}_{-0.04}$
for \rxte data
taken over a $\sim 3.2$~year period between 
1999--2002. Over such a long timescale variability can obviously
be a factor responsible for the smaller value of $R$.
 
Upon allowing the outer disk radius to be a free parameter in 
the original disk model, with $R$ fixed at 1.0,
we obtained a lower limit on its value of $\sim 540r_{g}$. 
The inner disk radius and $q$ are correlated in the sense that 
they affect the overall width of the emission line: larger inner
radii allow steeper radial emissivity index ($q$) since the
part of the disk giving the broadest part of the line with a steep
$q$ would then be missing. The data do not allow the ambiguity between
$q$ and the inner radius to be removed. By examining confidence
contours of $q$ versus $r_{\rm in}$, we found that a radial
emissivity steeper than even $r^{-3}$ is allowed if $r_{\rm in}$
is of the order of $100r_{g}$ or larger.

We note that our results for the relativistic disk line
are {\it not} inconsistent with the findings of Reeves \etal (2004),
namely that the \xmm data do not {\it require} a disk line when the
warm absorber is taken into account. Reeves \etal (2004) obtained
a 90\% upper limit on the EW of a disk line of 35~eV. However,
this assumed a steep emissivity index ($q=3$)
combined with a small inner disk radius ($r_{\rm in}=6r_{g}$), 
and a disk inclination
angle of $30^{\circ}$. With a flatter emissivity and a smaller
inclination angle, as obtained from the \chandra HEG data, 
a larger EW is allowed because the disk line width is narrower.
Thus, the \chandra HEG and \xmm measurements are consistent
with each other.
The reason why the \xmm data do not yield a significant detection of
the disk line is that the \chandra HEG data are more sensitive 
to a {\it narrow disk line}, due
to the factor of $\sim 4$ better spectral resolution. 
In \figcompshoulder(c) and \figcompshoulder(d), we show
that both the best-fitting \chandra HEG Compton-shoulder
model and disk-line model respectively, fit the \xmm data
equally well. For fitting the \xmm data (the same data
as in Reeves \etal 2004, which were
taken at a different time to the \chandra data), all
parameters except the overall normalization and the power-law
slope were fixed at their \chandra best-fitting values.
Again, the He-like Fe absorption feature discussed in \S\ref{heabs}
was deliberately not modeled here in order to facilitate
comparison between the \chandra and \xmm data.

Although Fe~$K\beta$ line emission is expected to accompany
\fekalfa line emission, the HEG data are not so sensitive to Fe~$K\beta$ 
lines because the branching ratio is 17:150 for $K\beta:K\alpha$.
Having said that, line-like residuals are apparent
in the spectral ratios shown in \figsixspecrat but some
part of this line emission could be due to \feklyap.
Reeves \etal (2004) reported significant line emission at
$\sim 7$~keV from \xmm data and obtained an EW of
$20 \pm 5$~eV for \feklya after accounting for \fekbeta emission.
Since Fe~$K\beta$ line emission is {\it expected} to
accompany the \fekalfa line emission, we self-consistently
added Gaussian and disk-line \fekbeta emission with the
expected branching ratio to the Gaussian plus disk-line 
model described above in order to test the
\chandra HEG data for \feklya emission. We modeled the
latter with an additional Gaussian, fixing the energy at
6.966~keV (e.g. see Pike \etal 1996) but leaving the
intrinsic width free for the summed, $\sim 830$~ks HEG spectrum,
but fixing it at the resulting best-fit value for the
individual spectra from observations (1)--(5).  
The best-fitting intrinsic width 
from the summed HEG spectrum  was
$9200 \rm \ km \ s^{-1}$ FWHM. However,
we did not obtain a very significant detection ($\Delta C$ was
always less than 2.7) 
for either the summed spectrum (90\%, one-parameter
upper limit: 17~eV) or the individual spectra from
the five observations. Although \figsixspecrat shows
evidence of variability of the line-like residuals
at $\sim 7$~keV, we could only obtain upper limits
on the EW ($29, 27, 65, 41$, and 14~eV for observations (1), (2), (3), (4)
and (5) respectively). All of the values of the \feklya EW upper
limits we obtained are consistent with the
EW measurements from \xmm data by Reeves \etal (2004).

\section{CONCLUSIONS}
\label{conc}

We have presented results of X-ray spectroscopy of the Fe-K line
region in \src from an extended monitoring
campaign with the \chandra High Energy Grating Transmission Spectrometer
(\hetgp). Consistent with previous studies, the \fekalfa line
core is resolved in the time-averaged spectrum
by the High Energy Grating (HEG) and has
a center energy indicating an origin in neutral or lowly-ionized Fe.
Despite a factor of $\sim 1.5$ variation in the X-ray continuum luminosity,
we measured no variability in the intensity of the \fekalfa line core
during five observations comprising the observing campaign, over $\sim 125$
days. The lack of response to the continuum is consistent with the
fact that 
a virial interpretation of the line FWHM ($\sim 1700 \ \rm km \ s^{-1}$)
gives a distance between the putative central black hole and the line emitter
of at least 0.06pc, or 70 light days. However, we detected marginal
evidence for variability in the intrinsic width
of the \fekalfa line core. In two out of
the five snapshots the \fekalfa line was resolved and
had a FWHM of $\sim 2300 \ \rm km \ s^{-1}$, 
whilst in the other three snapshots the
line was unresolved. This apparent variability was not related to
the X-ray continuum level. At 68\% confidence, the
\fekalfa line core was broader and had a higher flux in
the first two $\sim 170$~ks snapshots than in the
last three. If the line width variability is real it
could be due to changes in the matter distribution and/or
ionization state as a function of distance from the central engine
and would be important to investigate with future missions.

We found that the excess flux around the base of the \fekalfa line
core can be modeled by an emission line from a relativistic disk
rotating around a black hole, as well as a model
in which line core photons are
Compton-scattered 
(forming a ``Compton shoulder'' on the red side of the line peak).
More realistic modeling should self-consistently include
Compton-scattering of line photons, whether the line photons
originate in an optically-thick disk or more distant
matter. Also, the Compton-scattered continuum should be
self-consistently computed.
In principle, measurement of the unscattered and scattered line
photons, along with the reflected continuum, would constrain
whether the core of the \fekalfa line originates in optically-thick
matter (such as the putative obscuring torus)
or in optically-thin matter (e.g. the BLR and/or NLR).
However, the signal-to-noise of the current data
is not sufficient to distinguish between these scenarios.
We note that a Compton shoulder has
not been detected in any other \chandra \hetg observation
of a Seyfert~1 galaxy (Yaqoob \& Padmanabahn 2004). If this
is verified by higher signal-to-noise data, it would suggest
that the \fekalfa line core is in general likely to originate in
optically-thin matter.
In the case of NGC~3783, the EW
of the \fekalfa line core deduced from the composite Gaussian plus
disk-line model can easily be produced by optically-thin matter
with $N_{H} \sim 10^{23} \ \rm cm^{-2}$, covering less than
half the sky (e.g. see Yaqoob \etal 2001).

Modeled with a relativistic disk line,
the disk inclination angle was constrained to be less than $11^{\circ}$
and the emission line flux from the disk is comparable to that in
the \fekalfa line core. The disk-line modeling included a
continuum with a complex photoionized absorber and Compton
reflection modifying the intrinsic continuum. 
We showed that the \chandra HEG data are consistent with the
\xmm data in the sense that the same disk-line model can
account for both data sets, even with these continuum complexities.
Of course, with the current data it is not possible
to rule out more complicated scenarios 
involving, for example, a disk line modified by Compton scattering,
as mentioned above.
We also confirmed previous detections 
({\it HETGS}: Kaspi \etal 2002; {\it XMM-Newton}: Reeves \etal 2004)
of an absorption feature in the Fe-K band
(at $6.67 \pm 0.04$~keV, and with an EW of $17 \pm 5$~eV), 
likely
to be due to He-like Fe absorption.
On the other hand, evidence for \feklya emission was marginal in the HEG data.

Future observations with higher spectral resolution,
as afforded by \astroep, will be able to resolve important
ambiguities remaining in the Fe-K band in NGC 3783. In particular,
it will be possible to determine
if the \fekalfa line core is composed of more than one
line component and to determine if the broad part of the
\fekalfa line is really due to a relativistic disk.

The authors thank Ian George, Jane Turner, Barry McKernan, and
Shai Kaspi
for valuable discussions, and S. Watanabe for use of his Comptonized
line model.
T.Y. gratefully acknowledges support from
NASA grants NNG04GB78A, NAG5-10769, and AR4-5009X, the latter
issued by the Chandra X-ray Observatory Center,
which is operated by the Smithsonian Astrophysical Observatory for and
on behalf of NASA under contract NAS8-39073.
This research
made use of the HEASARC online data archive services, supported
by NASA/GSFC. This research has made use of the NASA/IPAC Extragalactic Database
(NED) which is operated by the Jet Propulsion Laboratory, California Institute
of Technology, under contract with NASA.
The authors are grateful to the \chandra 
instrument and operations teams for making these observations
possible.


\newpage


\begin{thebibliography}{59}


\bibitem{arnaud1996} Arnaud, K. A. 1996, Astronomical Data Analysis
Software and Systems V, eds. Jacoby, G., \& Barnes, J.,
ASP Conference Series, Vol. 101, p. 17

\bibitem{behar2003} Behar, E., Rasmussen, A. P., Blustin, A. J.,
Sako, M., Kahn, S. M., Kaastra, J. S., Branduardi-Raymont, G.,
\& Steenbrugge, K. C. 2003, ApJ, 598, 232

\bibitem{derosa2002}
De Rosa, A., Piro, L., Fiore, F., Grandi, P., Maraschi, L., Matt, G.,
Nicastro, F., \& Petrucci, P. O. 2002, A\&A, 387, 838

\bibitem{Fabi2000} Fabian, A. C.,  Iwasawa, K.,
 Reynolds, C. S., \& Young, A. J. 2000, PASP, 112, 1145

\bibitem{Fabi1989} Fabian, A. C., Rees, M. J., Stella, L., \&
White, N. E. 1989, MNRAS, 238, 729

\bibitem{Gabel2003a} Gabel, J. R. \etal 2003a, ApJ, 583, 178

\bibitem{Gabel2003b} Gabel, J. R. \etal 2003b, ApJ, 595, 120 

\bibitem{gehrels1986} Gehrels, N. 1986, ApJ, 303, 336

\bibitem{kaspi2001} Kaspi, S., \etal 2001, ApJ, 554, 216 

\bibitem{kaspi2002} Kaspi, S., \etal 2002, ApJ, 574, 643

\bibitem{krong2004} Krongold, Y., Nicastro, F., Elvis, M.,
Brickhouse, N. S., Liedahl D., \& Mathur, S., 2003, ApJ, 597, 832

\bibitem{lee2002} Lee, J. C., Iwasawa, K., Houck, J. C., Fabian, A. C.,
Marshall, H. L., \& Canizares, C. R. 2002, ApJ, 570, L47

\bibitem{long2004}
Longinotti, A. L., Nandra, K., Petrucci, P. O., \& O'Neill, P. M. 2004, MN, 355, 929

\bibitem{lz2001} Lubi\'{n}ski, P., \& Zdziarski, A. A. 2001, MNRAS, 323, L37 

\bibitem{mark1995} Markert, T. H., Canizares, C. R., 
Dewey, D., McGuirk, M., Pak, C., 
\& Shattenburg, M. L. 1995, Proc. SPIE, 2280, 168

\bibitem{magdz1995} Magdziarz, P., \& Zdziarski, A. A. 1995,
MN, 273, 837

\bibitem{marko2003}
Markowitz, A., Edelson, R., \& Vaughan, S. 2003, ApJ, 598, 935

\bibitem{marko2004} Markowitz, A., \& Edelson, R. 2004, ApJ,
617, 939

\bibitem{matt2002} Matt, G. 2002, MN, 337, 147

\bibitem{Nand1997a} Nandra, K.,  George, I. M., Mushotzky, R. F., 
Turner, T. J., \& Yaqoob, T. 1997, ApJ, 477, 602 

\bibitem{netzer1990} Netzer, H. 1990, in Active Galactic Nuclei,
ed. R. D. Blandford, H. Netzer, \& L. Woltjer (Berlin: Springer), 137

\bibitem{netzer2003} Netzer, H., \etal 2003, ApJ, 599, 933

\bibitem{pike1996} Pike, C. D., \etal 1996, ApJ, 464, 487

\bibitem{Perola2002} Perola, G. C., Matt, G., Cappi, M., Fiore, F.,
Guainazzi, M., Maraschi, L., Petrucci, P. O., \& Piro, L. 2002, A\&A, 389, 802

\bibitem{peter2004} Peterson, B. M. \etal 2004, ApJ, 613, 682
 
\bibitem{pettru2002} Petrucci, P. O., \etal 2002, A\&A, 388, L5

\bibitem{Reev2003} Reeves, J. N. 2003, in ASP Conf. Ser.,
Active Galactic Nuclei, from Central Engine to Host Galaxy, ed. S Collin,
F. Combes \& I. Shlosman, (San Francisco: ASP), Vol. 290, 35

\bibitem{Reev2004}Reeves, J. N., Nandra, K., George, I. M.,
 Pounds, K. A., Turner, T. J., Yaqoob, T. 2004, ApJ, 602, 648

\bibitem{Reyn2002} Reynolds, C. S., \& Nowak, M. A. 2003, Phys. Rep., 377, 389


\bibitem{Sulentic1998} Sulentic, J. W., Marziani, P., Zwitter, T.,
Calvani, M., \& Dultzin-Hacyan, D. 1998, ApJ, 501, 54

\bibitem{turn1999} Turner, T. J., George, I. M., Nandra, K.,
\& Turcan, D. 1999, ApJ, 524, 667

\bibitem{turn2004} Turner, T. J., Kramer, S. B., \& Reeves, J. N.
2004, ApJ, 603, 62

\bibitem{watan2003} Watanabe, S. \etal 2003, ApJ, 597, L37

\bibitem{Wgyy2001} Weaver. K. A., Gelbord, J., \& Yaqoob, T. 2001, ApJ, 
550, 261

\bibitem{Yaq2001} Yaqoob, T., George, I. M., Nandra, K., Turner, T. J., 
Serlemitsos, P. J., \& Mushotzky, R. F. 2001, ApJ, 546, 759

\bibitem{Yaq2003a} Yaqoob, T., George, I. M., Kallman, T. R.,
Padmanabhan, U., Weaver. K. A., \&
Turner, T. J. 2003a, ApJ, 596, 85 

\bibitem{Yaq2003b} Yaqoob, T., McKernan, B., Kramer, S. B., Crenshaw, D. M.,
Gabel, J. R., George, I. M., \& Turner, T. J. 2003b, ApJ, 582, 105 

\bibitem{yaqoo2002} Yaqoob, T., Padmanabhan, U., Dotani, T.,
\& Nandra, K. 2002, ApJ, 589, 487

\bibitem{Yaqoo2004} Yaqoob, T., \& Padmanabhan, U. 2004, ApJ, 604, 63

\end{thebibliography}
\end{document}